\documentclass[12pt,preprint]{aastex}

\shorttitle{DUST AT METAL-RICH HELIUM WHITE DWARFS}
\shortauthors{Kilic et al.}

\begin{document}

\title{Near-Infrared Constraints on the Presence of Warm Dust\\ 
at Metal-Rich, Helium Atmosphere White Dwarfs}

\author{Mukremin Kilic\altaffilmark{1,5}, 
J. Farihi\altaffilmark{2,3,4,5}, 
Atsuko Nitta\altaffilmark{2,5}, and 
S. K. Leggett\altaffilmark{2,5}}

\altaffiltext{1}{Department of Astronomy, 
			The Ohio State University, 
			Columbus, OH 43210; kilic@astronomy.ohio-state.edu}
\altaffiltext{2}{Gemini Observatory, 
			670 N. A'ohoku Place, 
			Hilo, HI 96720; anitta,sleggett@gemini.edu}
\altaffiltext{3}{Department of Physics \& Astronomy, 
			University of California, 
			430 Portola Plaza, 
			Los Angeles, CA 90095}
\altaffiltext{4}{Department of Physics \& Astronomy, 
			University of Leicester, 
			Leicester LE1 7RH, 
			UK; jf123@star.le.ac.uk}
\altaffiltext{5}{Visiting Astronomer at the Infrared Telescope Facility, 
			which is operated by the University of Hawaii under 
			Cooperative Agreement no. NCC 5-538 with the 
			National Aeronautics and Space Administration, 
			Office of Space Science, Planetary Astronomy Program.}

\begin{abstract}

Here, we present near-infrared spectroscopic observations of 15 helium atmosphere, 
metal-rich white dwarfs obtained at the NASA Infrared Telescope Facility.  While a 
connection has been demonstrated between the most highly polluted, hydrogen 
atmosphere white dwarfs and the presence of warm circumstellar dust and gas, their 
frequency at the helium atmosphere variety is poorly constrained.  None of our targets 
show excess near-infrared radiation consistent with warm orbiting material.  Adding 
these near-infrared constraints to previous near- and mid-infrared observations, the 
frequency of warm circumstellar material at metal-bearing white dwarfs is at least 20\% 
for hydrogen-dominated photospheres, but could be less than 5\% for those effectively 
composed of helium alone.  The lower occurrence of dust disks around helium atmosphere
white dwarfs is consistent with Myr timescales for photospheric metals in massive convection 
zones.  Analyzing the mass distribution of 10 white dwarfs with warm circumstellar material, 
we search for similar trends between the frequency of disks and the predicted frequency of 
massive planets around intermediate mass stars, but find the probability that disk-bearing 
white dwarfs are more massive than average is not significant.

\end{abstract}


\section{INTRODUCTION}

High gravity, in the absence of radiative forces, together with significant convection
zones, allows the downward sedimentation of heavy elements to proceed rapidly in 
cool white dwarfs (Schatzman 1958). Therefore, even if heavy elements are accreted, 
they will sink to the bottom of the photosphere quickly, typically leaving behind a near
pure hydrogen or helium atmosphere.  On the other hand, a recent survey of nearby 
cool DA (hydrogen-rich) white dwarfs by Zuckerman et al. (2003) found that up to 25\% 
of the isolated white dwarfs in their sample show trace amounts of heavy elements 
(mainly calcium, type DAZ)\footnote{Koester et al. (2005a) found a lower frequency of DAZs ($\approx$5\%)
among their DA white dwarf sample. Different biases contribute to the varying
percentages between the Zuckerman et al. (2003) and Koester et al. studies. Since the
biases for these two surveys have not been accounted for, a definitive
fraction of DAZ vs. DA has yet to be determined.}.
Since the diffusion timescales for heavy elements are 
shorter than the white dwarf evolutionary timescales, the source of the metals observed 
in the photospheres of these white dwarfs cannot be primordial.

The interstellar medium (ISM) seems like a natural candidate for the source of metals in 
white dwarf atmospheres. However, the Sun is located in a very low density region, and the 
volume filling factor of clouds in the local ISM is much less than 25\%, the percentage of DA
stars showing metallic features in the nearby white dwarf sample.  In addition, Zuckerman et 
al. (2003) and Kilic \& Redfield (2007) found no correlation between the accretion density 
required to supply metals observed in DAZ white dwarfs with the densities observed in their 
interstellar environment, indicating that ISM accretion alone cannot explain the presence of 
metals in these nearby degenerates.  Helium-rich white dwarfs with metals (type DBZ) do not 
show a correlation with the local ISM clouds, either (Aannestad et al. 1993).
Jura (2006) demonstrates there are several DBZ stars with carbon abundances at least an 
order of magnitude below solar, which is inconsistent with accretion from the ISM.

A more exciting possibility for the atmospheric metals is accretion from circumstellar disks. 
Until recently, there was only a single metal-rich white dwarf known to have a debris disk, G29-38 
(Graham et al. 1990; Zuckerman \& Becklin 1987; Koester et al. 1997).  Kilic et al. (2005) and 
Becklin et al. (2005) both independently discovered the second known debris disk around a 
metal-rich white dwarf star, GD 362.  Recently, 7 more dust disks have been discovered around 
metal-rich white dwarfs with space- and ground-based observations (Kilic et al. 2006; Kilic \& 
Redfield 2007; von Hippel et al. 2007; Jura et al. 2007; Farihi et al. 2008a), while no disks were 
indicated in sensitive space-based observations of $N\approx120$ non-metal-bearing white 
dwarfs (Mullally et al. 2007).

For the white dwarfs with debris disks, the presence of metals in the atmosphere is 
almost certainly related to the circumstellar material at these stars. If the source of the 
photospheric metals observed in several DAZ stars is accretion from circumstellar disks, 
there is no a priori reason to assume that the same mechanism does not work for DBZ 
white dwarfs.  It is perhaps noteworthy that the fraction of metal-bearing to non-metal-bearing 
stars is similar for both hydrogen and helium atmosphere white dwarfs (Koester et al. 2005a).  
However, because helium is relatively transparent compared to hydrogen, DB stars and their 
cooler counterparts have photospheres which are more readily contaminated than DA stars 
(Zuckerman et al. 2003). Additionally, these heavy contaminants can stay present in cool 
helium atmospheres for Myr timescales, owing to massive convection zones (Paquette et 
al.1986). Still, a similar mechanism may explain both the DAZ and DBZ stars.

Jura et al. (2007) obtained {\em Spitzer} IRAC and MIPS photometry of the carbon-deficient 
DBZ, GD 40 and found it to have a disk, while 2 other DBZ stars in the same program did not show 
any evidence for warm or cool dust. Moreover, high-resolution optical spectroscopy of GD 362 reveals 
it has a significant amount of helium, and a more accurate spectral type is DAZB (Zuckerman et al. 
2007; Koester et al. 2005b).  The Mullally et al. (2007) sample contains only 2 DBZ stars, neither 
of which show mid-infrared excess.  There has not yet been a systematic search for debris disks 
around these helium-rich, externally polluted stars.

We present near-infrared spectroscopic observations of 15 helium atmosphere, metal-rich 
white dwarfs selected from the literature (Dupuis et al. 1993; Koester et al. 2005a; Eisenstein 
et al. 2006).  The physical characteristics of our sample are presented in Table 1, our observations 
are discussed in \S 2, and the results and analysis of the spectroscopic data are presented in \S 3.

\section{OBSERVATIONS}

We used the 0.8--5.4 $\mu$m Medium-Resolution Spectrograph and Imager (SpeX; Rayner et al. 
2003) on the NASA Infrared Telescope Facility (IRTF) to perform near-infrared spectroscopy on 15 
helium atmosphere, metal-rich white dwarfs and 1 DAZ white dwarf, HE 0106$-$3253.  The prism mode 
was employed with a 0.5$\arcsec$ slit to produce spectra with a resolving power of 90--210 (average of 
150) over the $0.8-2.5\mu$m range.  Our observations were performed under conditions of thin cirrus 
and partly cloudy skies on 2007 March 5--6 and November 3--4.  To remove the dark current and the 
sky signal from the data, the observations were taken at two different positions on the slit separated by 
10$\arcsec$.  Internal calibration lamps (a 0.1W incandescent lamp and an argon lamp) were used for 
flat-fielding and wavelength calibration, respectively. In order to correct for telluric features and flux 
calibrate the spectra, nearby bright A0V stars were observed at an airmass similar to the target star 
observations. We used an IDL-based package, Spextool version 3.4 (Cushing et al. 2004), to reduce 
our data.

In addition to our IRTF observations, we obtained near-infrared spectra of another DAZ, 
WD 1150$-$153, on the United Kingdom Infra-Red Telescope (UKIRT).  Two separate sets of 
observations were performed on 2007 June 13 and 2008 January 9 in service mode (Service Program 
1659) using the UKIRT 1--5 $\mu$m Imager Spectrometer (UIST; Ramsay Howat et al. 2004) with the 
$HK$ grism and a $0.6''$ slit, yielding spectra over $1.4-2.5\mu$m with a resolving power of 400.  A 
nearby F0V standard, HD 96220, was observed for flux calibration and telluric correction.

\section{RESULTS}

Figure 1 presents the flux calibrated spectra of 9 DBZ stars, ordered by $T_{\rm eff}$, compared 
to the predicted photospheric fluxes for each star. Blackbody spectral energy distributions (SEDs) 
are suitable for warm helium-rich atmosphere white dwarfs, and are therefore used here.  Most of 
our targets have reliable 2MASS $J$-band photometry, while some also have $H$- and $K$-band 
data. The 2MASS photometry is shown as filled circles with error bars; the blackbody and observed 
spectra are normalized to match the 2MASS photometry in the $J$-band.  A typical telluric spectrum 
observed at the IRTF is presented in the top panel.  The features observed in several stars in the 
range $1.35 - 1.45 \mu$m and $1.80 - 2.05\mu$m are telluric correction residuals.

Figure 2 presents our IRTF spectra of 6 DBZ stars found in the SDSS along with their optical 
photometry and spectroscopy. The red lines show the appropriate blackbody SEDs.  Only 1 of 
these stars, J1247+4934, is hot enough to show helium lines and therefore classified DBZ. 
Optical helium lines disappear below $T_{\rm eff}\approx12000$ K in white dwarf atmospheres, 
and hence one can conclude that these 5 stars are cooler than this; they are classified either DZ or 
DZA.  Even though blackbody SEDs do not reproduce the observed spectra shortward of 0.5$\mu$m 
due to strong metal lines, they reproduce the near-infrared SEDs fairly well. Fitting their optical and 
near-infrared colors with blackbody colors, we find that 4 of the stars are consistent with $T_{\rm eff}
\approx$11500 K, and 1 of them, J1351+4253, with $T_{\rm eff}\approx$7500 K.

None of the 15 helium atmosphere, metal rich white dwarfs in our sample, including GD 40 (WD 
0300$-$013), show near-infrared excess. The 2MASS photometry of GD 40 is slightly brighter in the 
$K$-band than predicted by our spectroscopy.  Jura et al. (2007) discovered a mid-infrared excess 
around GD 40, and were able to model the observed excess with a flat disk extending from 20 to 30 
$R_*$. Their model predicts little or no excess in the $K$-band, at odds with the 2MASS $K$-band 
photometry, but consistent with our IRTF spectrum.  Figure 3 presents the near-infrared spectrum 
of the most metal-rich DAZ in the Koester \& Wilken (2006) sample, HE 0106$-$3253, compared 
to the model-predicted photospheric flux.  Perhaps surprisingly, it does not have an excess 
indicating a warm dust disk.

Figure 4 presents 3 near-infrared spectra of WD 1150$-$153 obtained in 2006 (Kilic \& Redfield
2007), 2007 and 2008 (this paper), which were taken to search for variability.  The timescale for 
Poynting-Robertson drag at the inner disk edge of such a $T_{\rm eff}\approx12000$ K DAZ is close 
to 30 days (von Hippel \& Thompson 2007), and if the process which replenishes the warmest dusty
material is not continuous, it could result in K-band flux variations.  WD 1150$-$153 is a ZZ Ceti variable 
with a dominant period of 250 s and an optical pulsation amplitude of 0.8\% (Gianninas et al. 2006; 
Koester \& Voss 2006); expected to be an order of magnitude smaller in the near-infrared. Normalizing 
each spectrum by the average flux between 1.44 and 1.82$\mu$m (the 2MASS $H$-band), the total 
$K$-band flux increased by 0.2\% in 2007, and decreased by 4.3\% in 2008, relative to the total flux 
observed in 2006.  However, this level of variability is well within a typical $3-5$\% photometric error 
for a single near-infrared obervation in clear weather conditions, which we did not have.  Therefore, 
it is likely that these differences are due to calibration errors caused by variations in sky brightness
and extinction due to non-photometric conditions.  Further photometric observations are needed to 
rule out possible changes in the disk.  Kilic \& Redfield (2007) reported 2 near-infrared spectroscopic 
observations of GD 362 separated by nearly 1 year, which did not reveal any variations. 

\section{DISCUSSION}

\subsection{Helium versus Hydrogen Atmospheres}

Like their DAZ counterparts, most DBZ white dwarfs do not show near-infrared excess due 
to dust disks. The only reported helium-rich white dwarf with a clear $K$-band detectable (disk) 
excess is GD 362, yet this star is also very hydrogen-rich and DAZ-like, both in its spectral 
(Gianninas et al. 2004), compositional ($\log$ (H/He) = $-1.1$; Zuckerman et al. 2007), and 
physical properties (B. Hansen 2008, private communication). In order to better constrain the 
fraction of dusty white dwarfs among spectral types DBZ and DAZ, we now consider all such 
stars with published IRTF and {\em Spitzer} observations. Figure 5 presents a plot of $T_{\rm eff}$ 
versus calcium abundance for 39 DAZ stars (Koester et al. 2005a, Zuckerman et al. 2003), 37 of which 
have now been observed at the IRTF and/or with {\em Spitzer}.  There are 7 observed stars which 
have debris disks, corresponding to an overall fraction of 18.9$^{+8.0}_{-4.8}$\%.
Due to small sample size, we used a binomial probability distribution to derive statistical uncertainties
for the frequency of disks. Since the probability distribution is not symmetric about its maximum value,
we report the range in frequency that delimits 68\% of the integrated probability function as error bars.
These error bars are equivalent to 1$\sigma$ limits for a Gaussian distribution (see the discussions
in McCarthy \& Zuckerman 2004 and Burgasser et al. 2003). If we include the DAZ analog GD 362, the fraction
of DAZ white dwarfs with disks goes up to 21.1$^{+7.9}_{-5.1}$\%. 
Only 1 nearly pure DB white dwarf with metals is known to 
have a dust disk, the recently reclassified DBAZ GD 40 ($\log$ (H/He) = $-$6.0; Voss et al. 2007),
corresponding to 1 of 19 stars in Table 1 with $\log$ (H/He) $< -4$ (abundances typical for DB
white dwarfs with hydrogen; Voss et al. 2007), or 5.3$^{+10.2}_{-1.7}$\%. If we limit our sample to ground-based
observations only, this value goes down to 2.6$^{+4.3}_{-2.6}$\% for hydrogen-poor white dwarfs and 14.3$^{+10.8}_{-4.6}$\% for 
hydrogen-rich white dwarfs. Dust disks appear less frequent among helium versus hydrogen 
atmosphere white dwarfs, and the difference appears to be significant at the 2$\sigma$ level.

If the observed difference in the frequency of dusty DAZ and DBZ stars is real, and the 
above analysis supports that conclusion, then this result is consistent with the difference in their
respective, typical diffusion timescales.  The diffusion timescales for metals are a few to several 
orders of magnitude longer in helium-rich atmospheres compared to hydrogen-rich analogs.  For 
a $0.6M_\odot$, 10000 K helium atmosphere white dwarf, the diffusion timescale for calcium is on 
the order of $10^5$ yr (Paquette et al. 1986), whereas it is only $10^2$ yr for a DAZ of the same 
temperature.  Therefore, a lower fraction of dust disks around helium-rich degenerates may imply 
the disk lifetimes are shorter than the metal diffusion timescales, a scientifically interesting constraint 
on these poorly understood compact disks.  While there may once have existed circumstellar dust 
around some or many polluted, helium-rich white dwarfs, these have either completely dissipated 
(become fully consumed by the star), or remain in gaseous phase and hence undetectable in the 
infrared. Any photospheric metal remnants should persist for Myr timescales in the relatively 
massive convection zones of these helium degenerates, and, in principle, could even be 
periodically replenished by the ISM.

Recently, Farihi et al. (2007) announced the discovery of a dust disk around another 
helium-rich white dwarf, GD 16, which also possesses significant amounts of atmospheric 
hydrogen. GD 16 is classified DAZB (as GD 362) with $\log$ (H/He) = $-2.89$ (Koester et al. 
2005b).  The presence of appreciable amount of hydrogen in a helium-dominated atmosphere can 
shorten the diffusion timescales for heavy elements significantly, presumably due to the
outermost layers being dominated by the lighter gas.  Both GD 362 and GD 16 probably 
behave more like DAZ stars in terms of their diffusion timescales (B. Hansen 2008, private 
communication).  Hence, the discovery of disks around these particular helium- and hydrogen-rich
stars may not be suprising, as the stars are more likely to be presently accreting in order that the
observed metal lines are present in their optical spectra. 

GD 40, the only nearly pure DB with a dust disk, also has trace amounts of hydrogen. However, 
the hydrogen abundance in GD 40 is lower by 3--5 orders of magnitude relative to GD 16 and GD 
362. Voss et al. (2007) found that 55\% of DB white dwarfs in the temperature range 10000 K -- 
30000 K show hydrogen at a similar level to GD 40 (log(H/He)$\la-4$).  This fraction goes up to 
62\% for the temperature range of our sample (10000 -- 20000 K).  It is therefore reasonable 
to assume that the source of hydrogen in GD 40 is also responsible for its presence in other 
DB white dwarfs (see Voss et al. 2007 for a thorough study and discussion of this poorly
understood phenomenon).

\subsection{Planet versus Disk Frequency}

Laws et al. (2003) and Fischer \& Valenti (2005), both suggest an increase in the incidence 
of planetary and substellar companions orbiting relatively more massive stars.  For planet masses 
$m>0.8\ M_{\rm J}$ and semimajor axes $a<2.5\ $AU, Johnson et al. (2007) find the frequency 
of planets increases from 4.2\% for $0.7M_\odot<M\leq1.3M_\odot$ stars, to 8.9\%, for $1.3
M_\odot<M\leq1.9M_\odot$ primaries.  The analogs of these planets around the progenitors of 
white dwarfs would almost certainly be engulfed on the AGB, and the frequency of planets around 
intermediate mass stars at wider separations is currently unknown. However, comparable numbers 
of planets may exist at wider separations. Kennedy \& Kenyon (2008) suggest a model in which the 
frequency of gas giant planets increases linearly with stellar mass from $0.4M_\odot$ to $3M_\odot$ 
and decreases for higher mass primaries. Recent surveys for planets around white dwarfs revealed a possible
planet around a white dwarf with $M_{\rm initial} \approx 2 M_\odot$ (Mullally et al. 2008) and no massive planets
around several dozen white dwarfs with $M_{\rm initial}\geq 3 M_\odot$ (Farihi et al. 2008b, see also
Burleigh et al. 2008; Friedrich et al. 2007; Debes et al. 2005, 2006).
If the white dwarf disks arise as a result of planetary 
system interactions, a correlation with degenerate stellar mass is possible.

von Hippel et al. (2007) studied the ensemble characteristics of 4 DAZ stars with debris disks. 
Including the recently discovered disk-bearing DAZ and DBZ stars, there are now 9 white dwarfs 
with dusty disks and 2 white dwarfs with gaseous disks (Gaensicke et al. 2007). Table 2 presents 
$T_{\rm eff}, \log$ g, initial and final masses for these white dwarfs, excepting GD 40 which has no
mass estimate available in the literature.  A typical error in these spectroscopic masses is around 
0.03$M_\odot$ (Liebert et al. 2005).  The initial masses for the progenitors of these white dwarfs are 
estimated using the relation given by Kalirai et al. (2008), ranging from $1.1M_\odot$ to $3.8M_\odot$ 
with an average of $2.3 M_\odot$. The use of an independent initial-to-final mass relation would 
change these mass estimates slightly. For example, employing the relation of Dobbie et al. (2006), Williams (2007), 
or Weidemann (2000) increases the initial mass estimates on average by 0.1 -- 0.3 $M_\odot$ compared 
to the Kalirai et al. (2008) relation.

Panel (a) of Figure 6 presents the mass distributions for the progenitors of the 10 white 
dwarfs presented in Table 2. The average and standard deviation for these disk-bearing white 
dwarf masses, including an assumed measurement error of 0.03$M_\odot$, is 0.65 $\pm$ 0.09 
$M_\odot$.  Although slightly higher, this value is consistent within the errors with the mean mass 
for the field DA (0.60 $\pm$ 0.13 $M_\odot$; Liebert et al. 2005) and DB (0.60 $\pm$ 0.07 $M_\odot$ 
(Voss et al. 2007) white dwarf samples.  One caveat, however, is the outstanding problem of DA white 
dwarf mass estimates from spectroscopy for effective temperatures less than 12000 K (for a thorough 
review of this problem, see Kepler et al. 2007).  In a nutshell, the presence of unseen helium (and possibly 
other neutral particles) can skew cooler DA masses towards higher values; perfectly exemplified by the 
mass estimate for GD 362 prior to the high resolution detection of atmospheric helium (Zuckerman et 
al. 2007; Gianninas et al. 2004).  There are only a few targets in Table 2 which belong in this category, 
and our results are unlikely to be affected by the situation, but we mention it for completeness and due
to the fact that DA field white dwarf mass measurements purposefully exclude stars in this temperature
range (Liebert et al. 2005).  Therefore, we cannot say whether a trend exists between disk frequency
and white dwarf mass, but a larger sample of such stars, combined with parallaxes to remove any mass
estimate ambiguity introduced via spectroscopy for cooler DA type stars, may prove more insightful on 
this astrophyscially important topic.  Based on previous works on white dwarfs with disks, it appears
probable that the circumstellar material should be associated with surviving planetary system bodies,
and hence any correlation between disks and host mass would imply a correlation between planetary
systems and host mass.

The cumulative mass distributions for hydrogen- and metal-rich white dwarfs with (DAZd) and 
without (DAZ) disks from the Koester \& Wilken (2006) analysis are shown in panel (b) of Figure 6. 
A Kolmogorov-Smirnov (K-S) test indicates that there is a 48\% probability that the DAZd white dwarfs 
are randomly drawn from the general distribution of DAZ white dwarfs. About half of these DAZ stars 
are also analyzed by P. Bergeron in various papers (e.g. Liebert et al. 2005; Bergeron et al. 2001); 
panel (c) shows the DAZ mass distribution using these mass estimates. A K-S test shows that the 
probability that the DAZ and DAZd distributions are similar is 81\%. Thus, based on the 7 DAZd white 
dwarfs, there is no correlation apparent between the frequency of disks and degenerate mass.  Panel 
(d) of Figure 6 shows the cumulative mass distributions for the DA and DAZ white dwarfs analyzed by
Zuckerman et al. (2003).  Masses for these white dwarfs are derived from the quoted stellar parameters, 
where known (i.e. log $g=8$ not assumed), and Bergeron et al. (1995) tabulated models. A K-S test 
shows that the null hypothesis is 28\% probable; we do not find a significant difference between the 
DA and DAZ mass distributions.

If all metal-rich white dwarfs accrete from dusty or gaseous circumstellar material, and if there is a 
connection between white dwarf disks and planets, then we might expect metal-rich white dwarfs to 
be more massive than normal white dwarfs if planets occur more frequently around intermediate 
mass stars.  However, there exists theoretical and now some empirical evidence that certain types
of planetary systems may induce significant mass loss on the first and asymptotic giant branches 
(Siess \& Livio 1999a,b).  Nelemans et al. (1998) proposed a scenario to explain the formation of 
single, low mass white dwarfs via enhanced mass loss due to planets or brown dwarfs which spiral
into the first ascent giant envelope; essentially analogous to the standard (binary) formation scenario 
for low mass, helium core white dwarfs and subdwarf B stars (Han et al. 2002).  Moreover, the 3 $M_J$ 
planet orbiting the sdB star V 391 Pegasi (Silvotti et al. 2007) may have belonged to a multiple planet 
system in which the innermost planet(s) were cannibalized to create the 0.5 $M_\odot$ helium-burning 
primary, which should evolve directly into a low mass white dwarf.  Therefore, any correlation between 
stellar mass and planet-disk frequency could be erased or marginalized by similar processes.

\section{Conclusions}

Our search for near-infrared excess at 15 metal-rich, helium atmosphere white dwarfs did not 
reveal any new evidence for disks.  We tentatively conclude that circumstellar material is responsible 
for the photospheric metals in at least 20\% of the hydrogen-rich white dwarfs and no more than 5\% of 
the hydrogen-poor white dwarfs.  The lower frequency of disks at externally polluted, helium-rich yet
hydrogen-poor white dwarfs possibly reflects a typical disk lifetime which is at least an order of magnitude
shorter than the $10^4-10^6$ yr diffusion timescales for metals in pure helium atmospheres.  On the other 
hand, the higher discovery rate of disks around metal-contaminated, hydrogen-rich white dwarfs (including 
types DAZ and DAZB) analogously reflects short diffusion times compared to disk lifetimes, which favor
discovery of disks in metal line DA white dwarfs.

In addition, we searched for $K$-band flux variations in the DAZ WD 1150$-$153, and did not observe
significant changes within 1.7 yr, implying that the warmest emitting material at the inner edge of the disk 
remains largely unchanged.  We also studied the mass distribution of metal-rich white dwarfs with disks in order 
to search for similar trends between the frequency of disks and the predicted frequency of gas giants 
around intermediate mass stars, but given the current small number statistics and errors, the probability 
that dusty white dwarfs are more massive than normal white dwarfs is not significant. 

\acknowledgements
MK and JF are grateful to B. Hansen for his insight on the properties of mixed atmosphere 
white dwarfs. MK thanks T. von Hippel and A. Gould for a careful reading of an earlier version 
of this manuscript. SKL's \& AN's research is supported by the Gemini Observatory, which is
operated by the Association of Universities for Research in Astronomy,
Inc., on behalf of the international Gemini partnership of Argentina,
Australia, Brazil, Canada, Chile, the United Kingdom, and the United
States of America. Some of the data reported here were obtained as part of the UKIRT Service 
Programme 1659. The UKIRT is operated by the Joint Astronomy Centre on behalf of the Science 
and Technology Facilities Council of the U.K. This publication makes use of data products from 
the Two Micron All Sky Survey, which is a joint project of the University of Massachusetts and the 
Infrared Processing and Analysis Center/California Institute of Technology, funded by the 
National Aeronautics and Space Administration and the National Science Foundation.

\clearpage
\begin{deluxetable}{lcccclll}
\tabletypesize{\footnotesize}
\tablecolumns{8}
\tablewidth{0pt}
\tablecaption{Metal-Rich, Helium Atmosphere White Dwarfs Observed in the Infrared}
\tablehead{
\colhead{Object}&
\colhead{$T_{\rm eff}$(K)}&
\colhead{log $g$}&
\colhead{$\log$ (Ca/He)}&
\colhead{$\log$ (H/He)}&
\colhead{Spectral Type}&
\colhead{Source}&
\colhead{Reference}
}
\startdata
WD 0002+729				& 13750 	& 8.00 	& $-$9.6 		& $-$6.0 		& DBZ 	& {\em Spitzer} 		& W02\\
WD 0300$-$013 (GD 40)		& 15200	& 8.00 	& $-$6.7 		& $-$6.0 		& DBZA 	& IRTF/{\em Spitzer} 	& F99, V07\\
HE 0449$-$2554 			& 12200 	& 8.20 	& $-$7.2 		& \nodata 		& DBAZ 	& IRTF 			& F00\\
J080537.64+383212.4 		& 10470 	& 8.82 	& \nodata 		& \nodata 		& DZ 	& IRTF 			& E06\\
		         				& 10660 	& 8.00 	& $-$10.03	& $<-$6.0 		& 		& 				& D07\\
J093942.29+555048.7 		& 10090 	& 7.51 	& \nodata 		& \nodata		& DZA 	& IRTF 			&E06\\
	                 				& 8680 	& 8.00 	& $-$8.51 		& $-$4.20 		& 		& 				& D07\\
J103809.19$-$003622.5 		& 10080 	& 7.28 	& \nodata 		& \nodata 		& DZ 	& IRTF 			&E06\\
			 			& 6770  	& 8.00 	& $-$9.44 		& $<-$5.0 		& 		& 				& D07\\
WD 1225$-$079	 		& 10500 	& 8.00 	& $-$7.9 		& \nodata 		& DZAB 	& IRTF 			& W02\\
J124703.28+493423.6 		& 16810 	& 8.21 	& \nodata 		& \nodata 		& DBAZ 	& IRTF 			&E06\\
J131336.95+573800.5 		& 10090 	& 8.82 	& \nodata 		& \nodata 		& DZ 	& IRTF 			&E06\\
			 			& 8900 	& 8.00 	& $-$9.39 		& $-$4.70 		& 		& 				& D07\\
HE 1349$-$2305 			& 18300 	& 8.13 	& $-$8.0 		& $-$4.9 		& DBAZ 	& IRTF 			& K05\\
J135118.47+425316.0 		& 10080 	& 8.82 	& \nodata 		& \nodata		& DZ 	& IRTF 			& E06\\
			 			& 6770 	& 8.00 	& $-$11.35 	& $<-$3.0 		& 	& 			& D07\\
WD 1352+004 				& 15300 	& 8.48 	& $-$9.3 		& $-$4.8 		& DBAZ 	& IRTF 			& K05\\
WD 1425+540 				& 14100 	& \nodata & $-$10.3 		& $-$4.2		& DBAZ 	& IRTF 			& D93\\
WD 1709+230 				& 19700 	& 8.17 	&  $-$8.0 		& $-$4.2 		& DBAZ 	& IRTF 			& K05\\
WD 1729+371 (GD 362) 		& 10540 	& 8.24 	& $-$6.24 		& $-$1.14 		& DAZB 	& IRTF/{\em Spitzer} 	& Z07\\
WD 1822+410 				& 17000 	& 7.90 	& $-$8.15 		& $-$4.0 		& DBAZ 	& {\em Spitzer} 		& W02\\
WD 2142$-$169 			& 15900 	& 7.93 	& $-$9.1 		& $-$4.5 		& DBAZ 	& IRTF 			& K05\\
WD 2144$-$079 			& 16500	& 7.90 	& $-$8.5 		& \nodata 		& DBZ 	& {\em Spitzer} 		& K05\\
WD 2322+118 				& 12000	&  \nodata & $-$8.7		& $-$5.2		& DZA 	& IRTF 			& D93\\
WD 2354+159 				& 19000	& 7.91 	& $-$8.1 		& \nodata 		& DBZ 	& {\em Spitzer} 		& K05\\	
\enddata
\tablecomments{The IRTF and {\em Spitzer} observations are from: this study; Kilic et al. (2005); 
Mullally et al. (2007); and Jura et al. (2007). The IRTF observations are sensitive for excess emission 
from the disks in the $H$- and $K$-bands, whereas the {\em Spitzer} observations are sensitive 
up to 8$\mu$m.}

\tablerefs{
D93 (Dupuis et al. 1993); 
D07 (Dufour et al. 2007); 
E06 (Eisenstein et al. 2006); 
F99 (Friedrich et al. 1999); 
F00 (Friedrich et al. 2000); 
K05 (Koester et al. 2005a); 
V07 (Voss et al. 2007); 
W02 (Wolff et al. 2002); and 
Z07 (Zuckerman et al. 2007).}
\end{deluxetable}

\clearpage
\begin{deluxetable}{lccccl}
\tabletypesize{\footnotesize}
\tablecolumns{6}
\tablewidth{0pt}
\tablecaption{Possible Progenitor Masses for White Dwarfs with Circumstellar Disks}
\tablehead{
\colhead{Object}&
\colhead{$T_{\rm eff}$(K)}&
\colhead{log $g$}&
\colhead{$M_{\rm final}(M_\odot)$}&
\colhead{$M_{\rm initial}(M_\odot)$}&
\colhead{Reference}
}
\startdata
WD 0408$-$041 	&    15070 	& 7.96 	&  0.59 	&      1.8 	& K01\\
WD 1015+161 		&    19540 	& 8.04 	&  0.65 	&      2.3 	& L05\\
WD 1116+026 		&    12290 	& 8.05 	&  0.63 	&      2.2 	& L05\\
WD 1150$-$153 	&    12260 	& 7.83 	&  0.51 	&      1.1 	& K01\\
WD 1455+298 		&    7390  		& 7.97 	&  0.58 	&      1.7 	& L05\\
WD 2115$-$560 	&    9940  		& 8.13 	&  0.66 	&      2.4 	& B95\\
WD 2326+049 		&    11820 	& 8.15 	&  0.70 	&      2.8 	& L05\\
GD 362      		&    10540 	& 8.24 	&  0.74 	&      3.2 	& Z07\\
J1043+0855 		&    18330 	& 8.09 	&  0.67 	&      2.5 	& G07\\
J1228+1040 		&    22290 	& 8.29 	&  0.81 	&      3.8 	& G07\\
\enddata
\tablecomments{
References given are for the white dwarf mass, as determined
primarily via Balmer line spectroscopy.}

\tablerefs{
K01 (Koester et al. 2001); 
L05 (Liebert et al. 2005); 
B95 (Bragaglia et al. 1995); 
Z07 (Zuckerman et al. 2007); and 
G07 (Gaensicke et al. 2007).}
\end{deluxetable}

\clearpage
\begin{figure}
\plotone{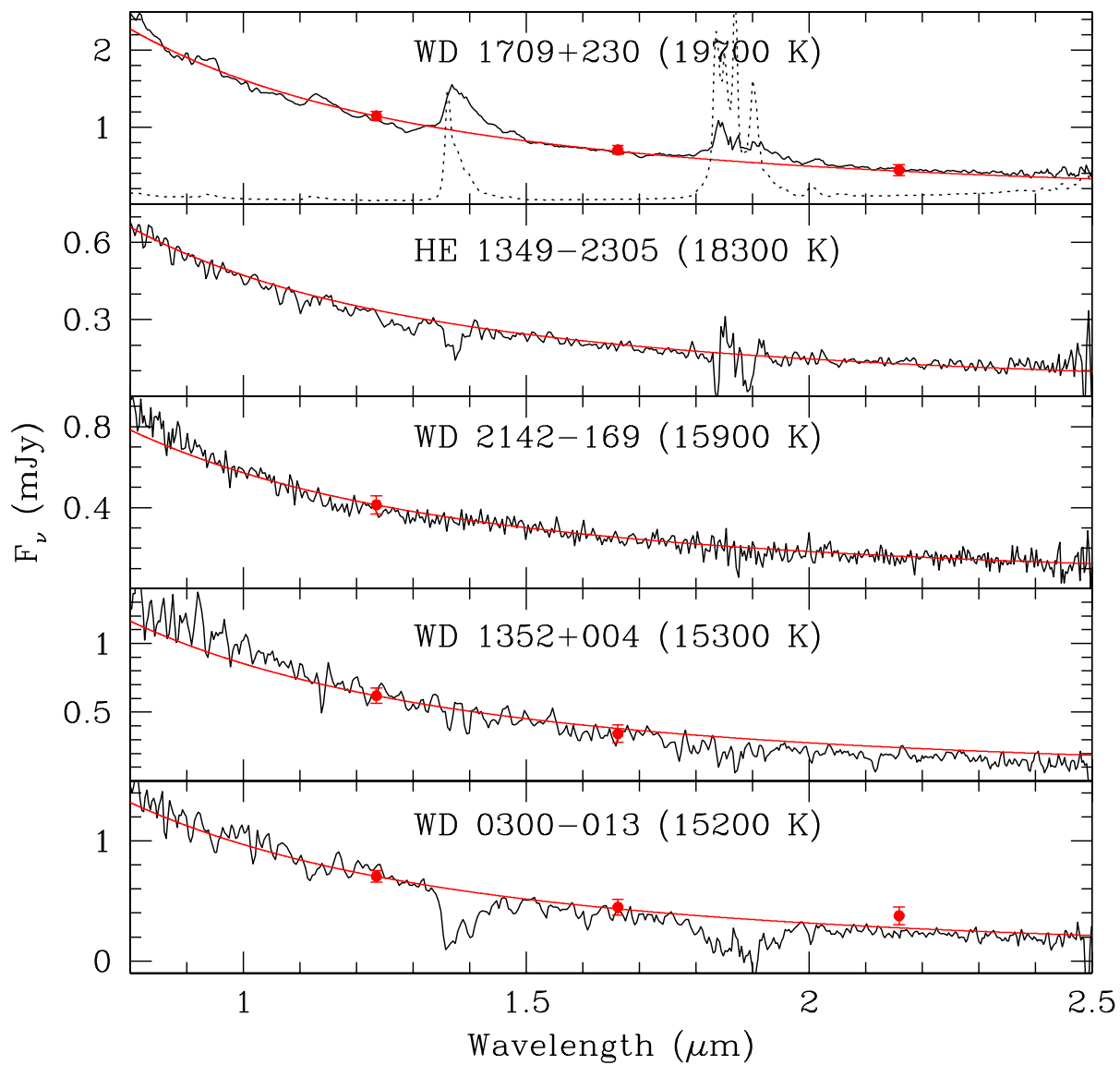}
\caption{Flux calibrated spectra of the helium atmosphere white dwarfs in our 
sample (black lines, ordered in $T_{\rm eff}$) compared to models (red lines). 
The 2MASS photometry is shown as filled circles with error bars. Telluric 
spectrum obtained from observations of WD 1709+230 are shown (dotted line) in the 
top panel.}
\end{figure}

\clearpage
\begin{figure}
\figurenum{1}
\plotone{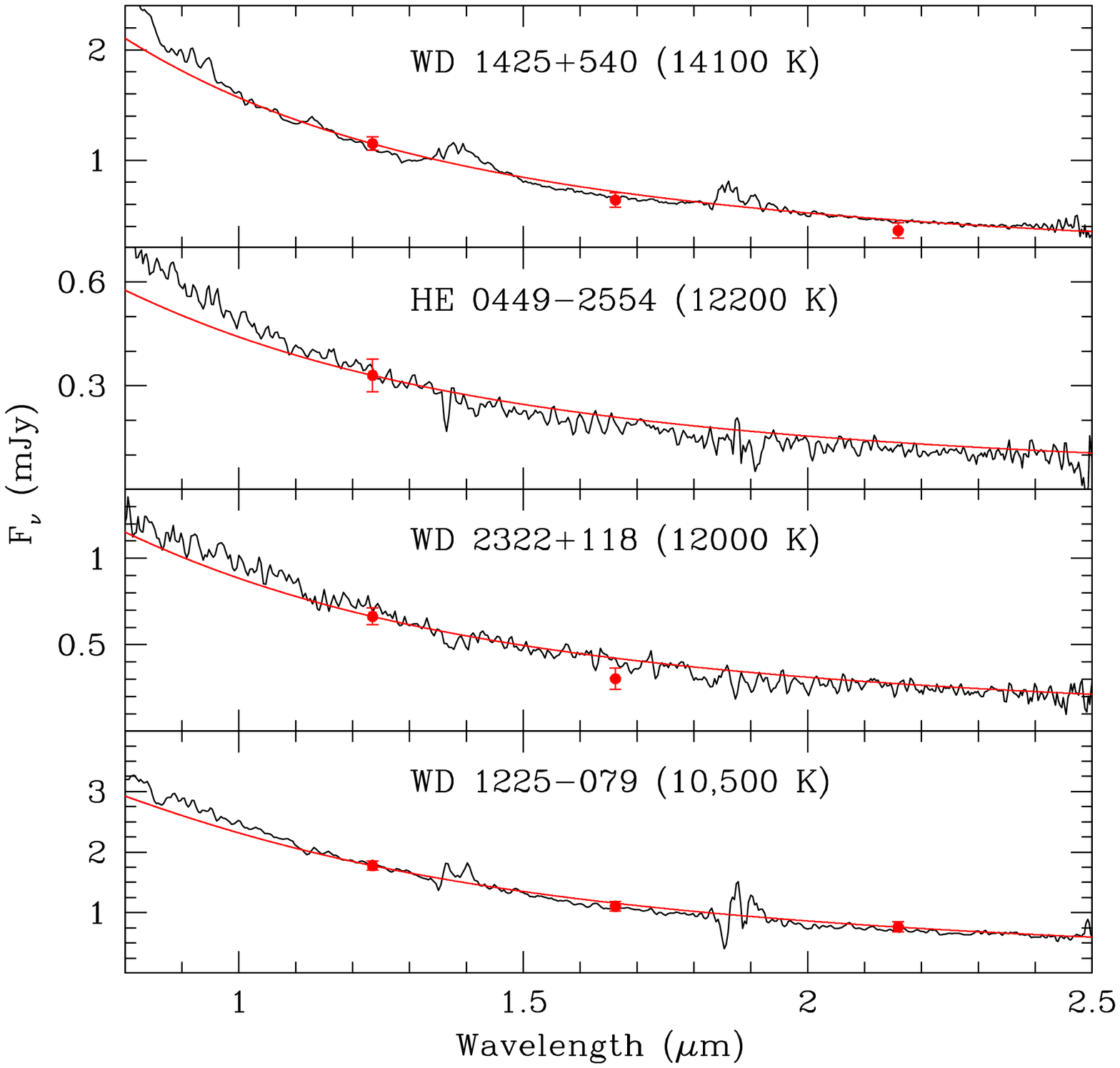}
\caption{cont.}
\end{figure}

\clearpage
\begin{figure}
\plotone{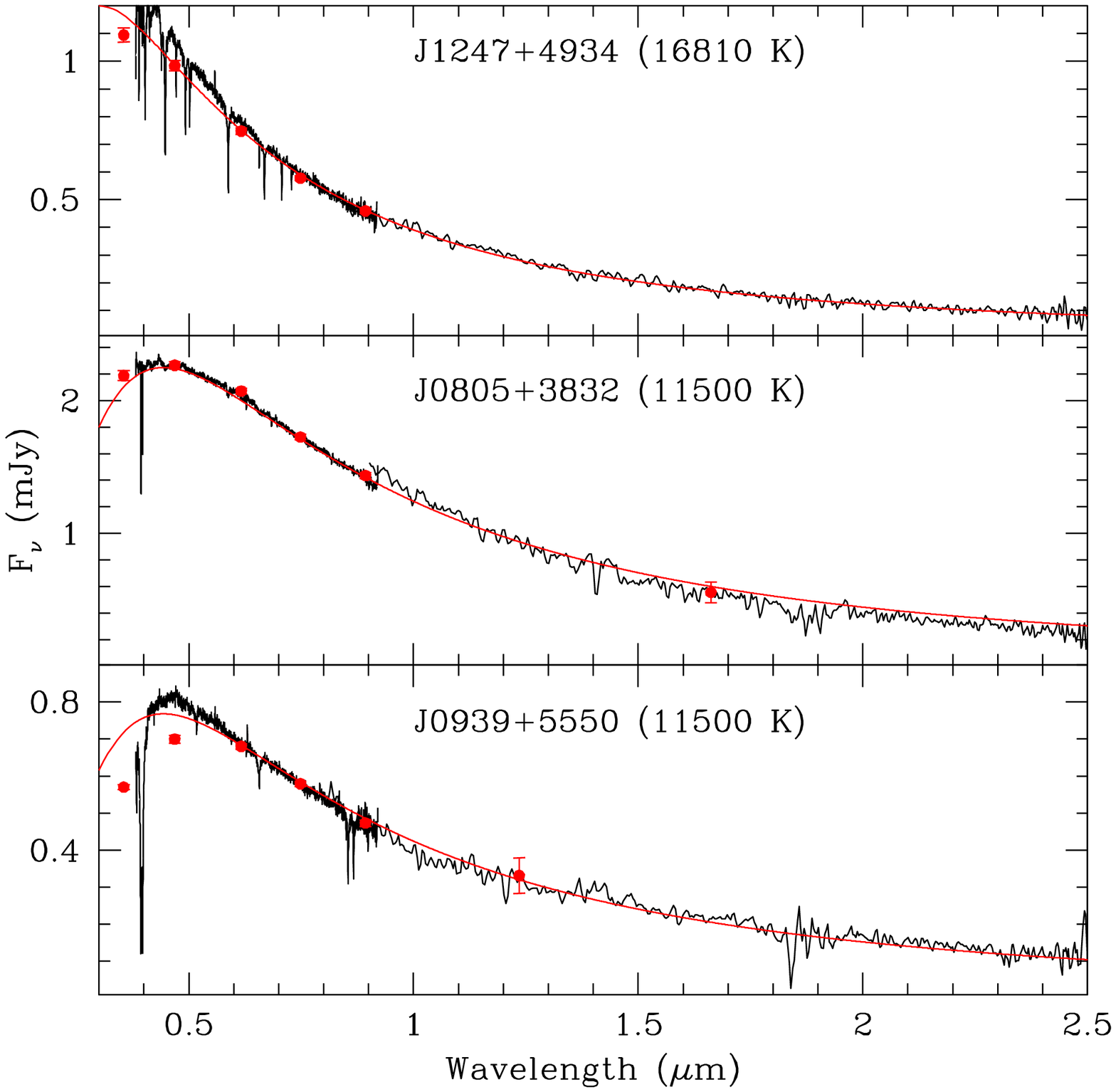}
\caption{Flux calibrated spectra of the SDSS DBZ and DZ white dwarfs (black lines) compared 
to models (red lines). The SDSS and 2MASS photometry are shown as filled circles with error bars.}
\end{figure}

\clearpage
\begin{figure}
\figurenum{2}
\plotone{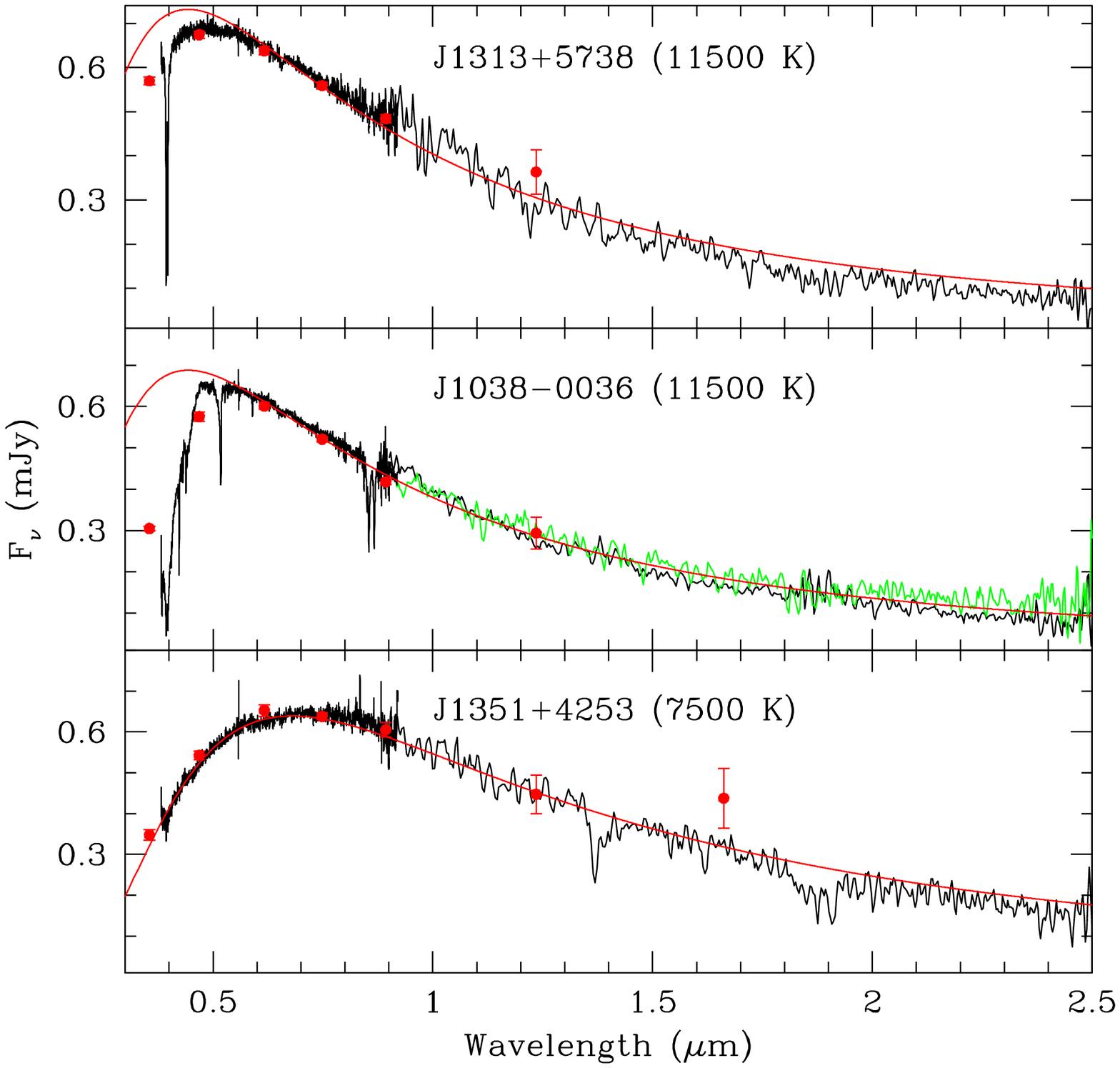}
\caption{cont.}
\end{figure}

\clearpage
\begin{figure}
\includegraphics[scale=0.65,angle=-90]{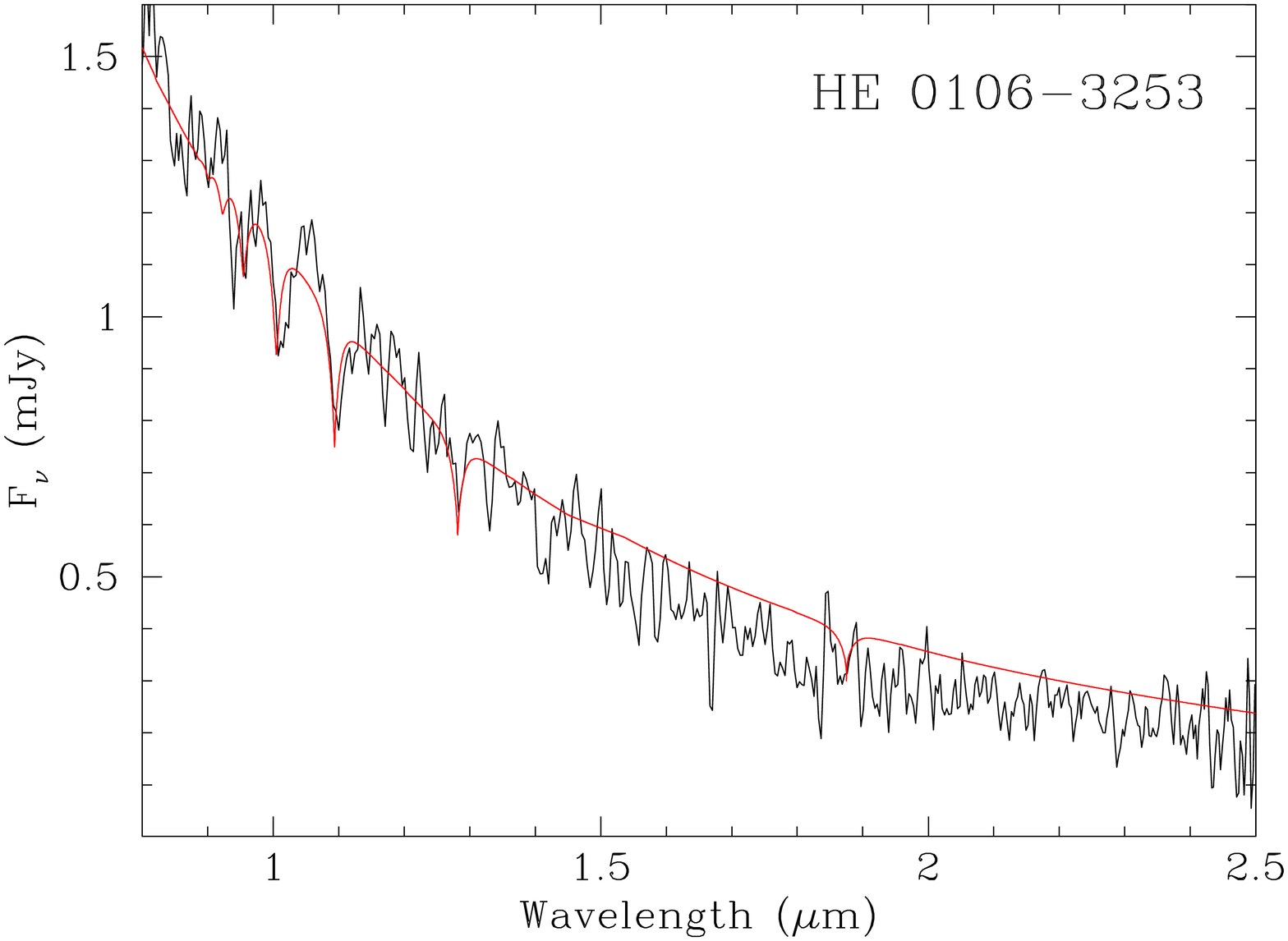}
\caption{Near-infrared spectrum of HE 0106$-$3253 (black line) compared to a 
$T_{\rm eff}=15700$ K DA white dwarf model kindly provided by D. Koester (2007,
private communication, red line).}
\end{figure}

\clearpage
\begin{figure}
\plotone{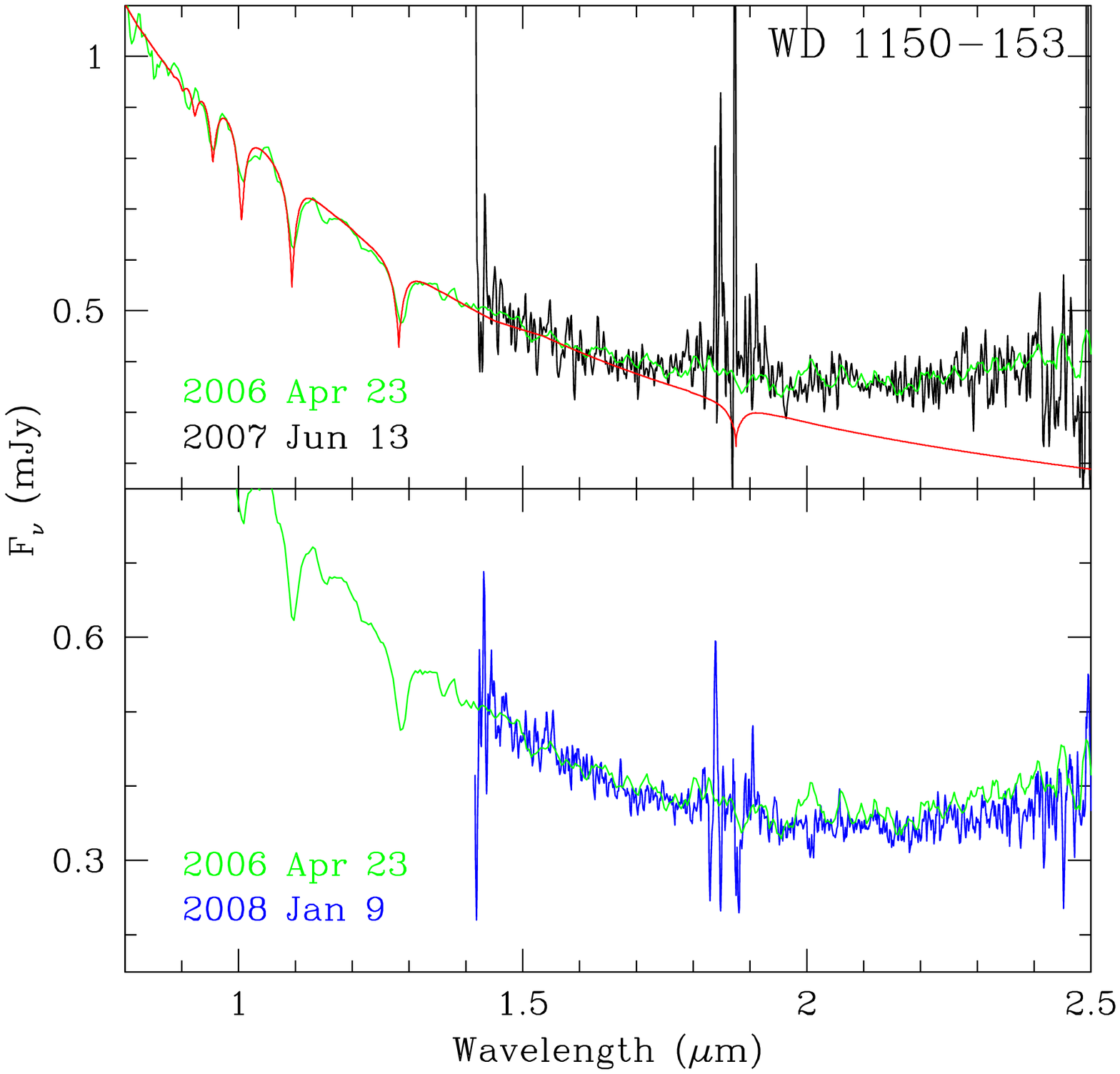}
\caption{Near-infrared spectra of WD 1150$-$153 obtained on 2006 April 23 (IRTF, green line),
2007 June 13 (UKIRT, black line), and 2008 January 9 (UKIRT, blue line). The red line shows
the expected photospheric flux from the white dwarf.}
\end{figure}

\clearpage
\begin{figure}
\plotone{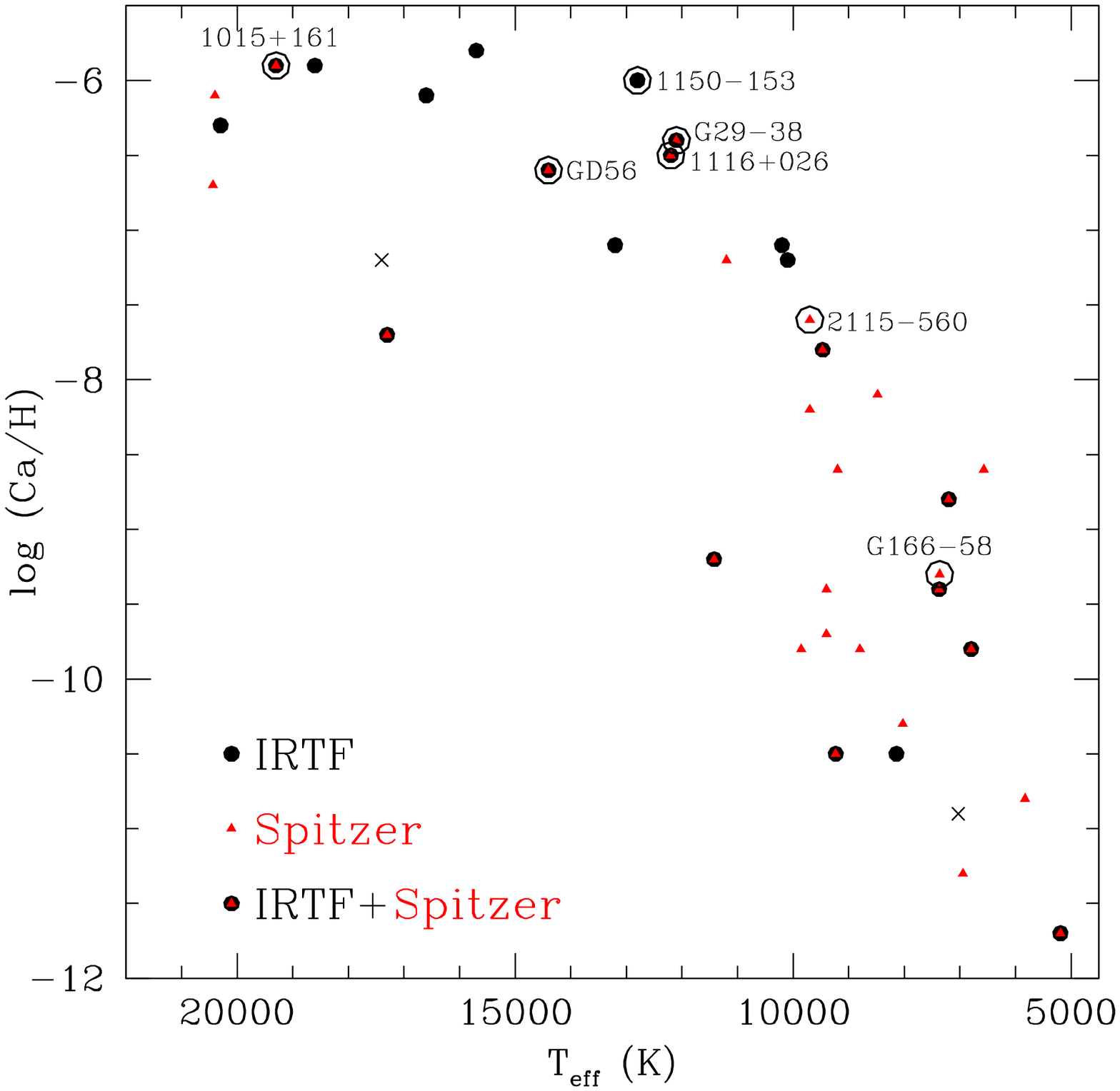}
\caption{Calcium abundance versus effective temperature for the DAZ stars observed 
at the IRTF (filled circles; Kilic et al. 2006, Kilic \& Redfield 2007, this study) and
{\em Spitzer}/IRAC (filled triangles; von Hippel et al. 2007, Debes et al. 2007, Farihi et al. 2008a). 
The remaining DAZ white dwarfs from Koester \& Wilken (2006) are shown as crosses. White dwarfs with circumstellar 
debris disks are labeled and marked with open circles.}
\end{figure}

\clearpage
\begin{figure}
\plotone{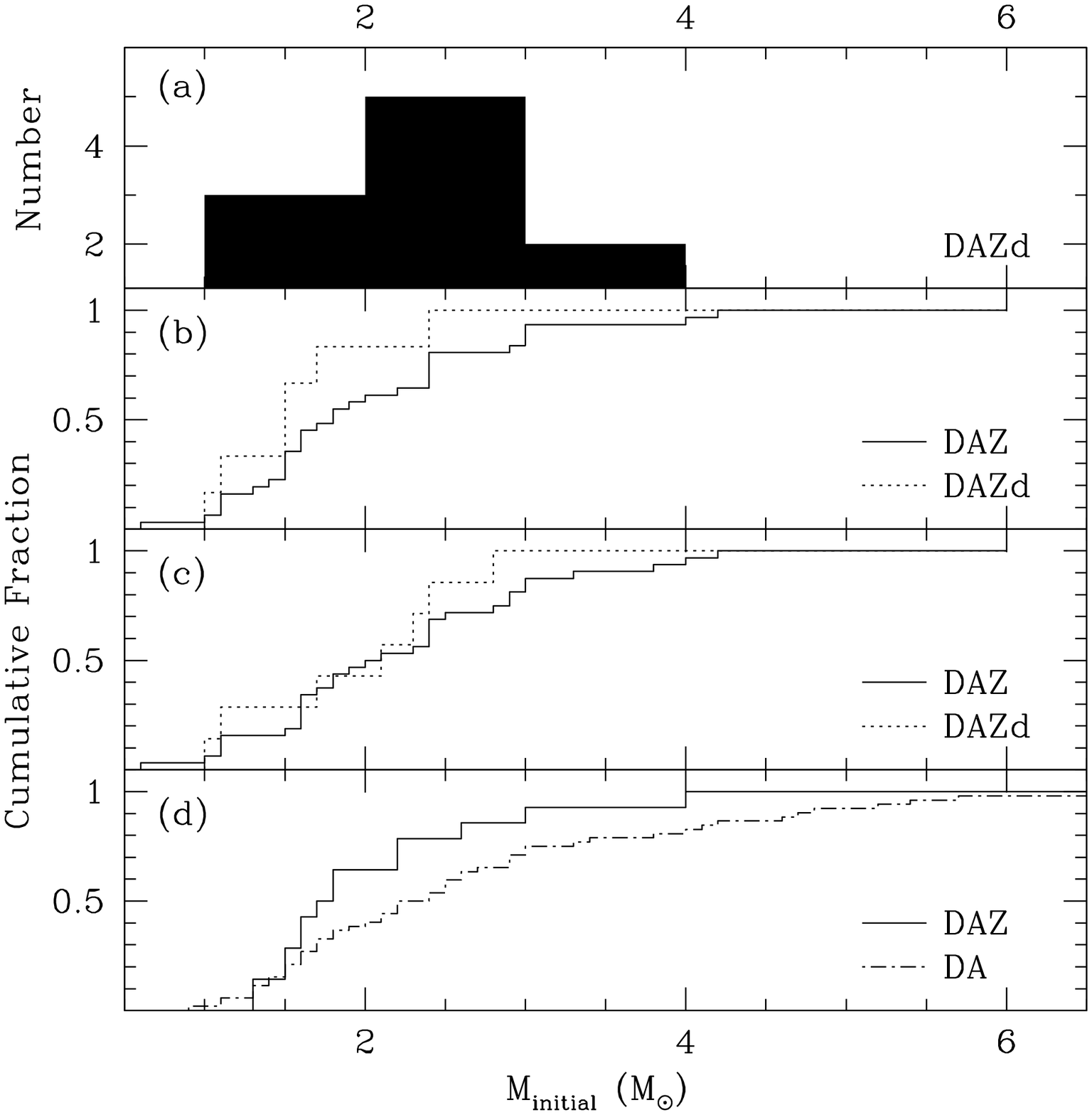}
\caption{Panel (a): the mass distribution for the progenitors of 10 white dwarfs with 
circumstellar disks.  Panels (b,c,d): the cumulative mass distributions for the DA, DAZ, and 
DAZd white dwarfs analyzed by Koester \& Wilken (2006), Bergeron et al. (2001), and 
Zuckerman et al. (2003), respectively.}
\end{figure}

\end{document}